\documentclass[10pt,twocolumn]{article}
\usepackage[top=2cm, bottom=2cm, left=1.8cm, right=1.8cm]{geometry}
\usepackage{times}  %
\usepackage[hyphens]{url}
\usepackage{graphicx}  %
\usepackage[keeplastbox]{flushend}
\usepackage{xcolor}
\usepackage{booktabs}
\usepackage{graphicx}
\usepackage{paralist}
\usepackage[small,bf]{caption}
\usepackage[tight]{subfigure}
\usepackage[hang,flushmargin]{footmisc}

\usepackage[compact]{titlesec}
\titlespacing*{\section}{0pt}{*4}{4pt}
\titlespacing{\subsection}{0pt}{*3}{3pt}
\usepackage{xspace}

\usepackage[hang,flushmargin]{footmisc}

\usepackage[compact]{titlesec}
\titlespacing*{\section}{0pt}{*4}{4pt} 
\titlespacing{\subsection}{0pt}{*3}{3pt}

\usepackage{enumitem}
\setlist[itemize]{noitemsep, topsep=0pt}
\setlist[enumerate]{noitemsep, topsep=0pt}
\setlist[description]{noitemsep, topsep=0pt}

\usepackage{multirow}
\usepackage{amsmath,bm,amssymb}
\usepackage{calc}
\usepackage{wasysym}

\usepackage[bookmarks=true, bookmarksnumbered=true, colorlinks=true, linkcolor=linkcol, citecolor=citecol, urlcolor=urlcol, hypertexnames=true]{hyperref}

\definecolor{darkgreen}{RGB}{0, 100, 0}
\definecolor{linkcol}{rgb}{0.3,0,0}
\definecolor{citecol}{rgb}{0.3,0,0}
\definecolor{urlcol}{rgb}{0.3,0,0}

\makeatletter
\def\url@leostyle{%
  \@ifundefined{selectfont}{\def\UrlFont{\small}}%
  {\def\UrlFont{}}%
}
\makeatother
\usepackage{ifthen}
\newif\ifshort
\shorttrue   %
\ifshort
  \newcommand{\isShort}{true}
\else
  \newcommand{\isShort}{false}
\fi
\newcommand{\shortVer}[1]{\ifthenelse{\equal{\isShort}{true}}{{#1}}{}}
\newcommand{\longVer}[1]{\ifthenelse{\equal{\isShort}{false}}{{#1}}{}}

\newif\ifcomment
\commenttrue
\ifcomment
\newcommand{\sz}[1]{{\bf \textcolor{blue}{SZ: #1}}}
\newcommand{\jbnote}[1]{{\bf \textcolor{magenta}{JB: #1}}}
\newcommand{\jf}[1]{{\bf \textcolor{red}{JF: #1}}}

\else
\newcommand{\sz}[1]{}
\newcommand{\jbnote}[1]{}
\newcommand{\jf}[1]{}
\fi

\newcommand{\descr}[1]{\smallskip\noindent\textbf{#1}}

\begin{document}
\title{\bf{The Pushshift Telegram Dataset}}

\author{\bf Jason Baumgartner\textsuperscript{\rm 1,*}, Savvas Zannettou\textsuperscript{\rm 2,\smiley{}}, Megan Squire\textsuperscript{\rm 3}, Jeremy Blackburn\textsuperscript{\rm 4,\smiley{}}\\[0.5ex] %
\normalsize \textsuperscript{\rm 1}Pushshift.io, \textsuperscript{\rm 2}Max Plank Institute, \textsuperscript{\rm 3}Elon University, \textsuperscript{\rm 4}Binghamton University\\
\normalsize \textsuperscript{*}Network Contagion Research Institute, \textsuperscript{\smiley{}}iDRAMA Lab\\
\normalsize jason@pushshift.io, szannett@mpi-inf.mpg.de, msquire@elon.edu, blackburn@cs.binghamton.edu
}
\date{}

\maketitle

\begin{abstract}
Messaging platforms, especially those with a mobile focus, have become increasingly ubiquitous in society.
These mobile messaging platforms can have deceivingly large user bases, and in addition to being a way for people to stay in touch, are often used to organize social movements, as well as a place for extremists and other ne'er-do-well to congregate.

In this paper, we present a dataset from one such mobile messaging platform: Telegram.
Our dataset is made up of over 27.8K channels and 317M messages from 2.2M unique users.
To the best of our knowledge, our dataset comprises the largest and most complete of its kind.
In addition to the raw data, we also provide the source code used to collect it, allowing researchers to run their own data collection instance.
We believe the Pushshift Telegram dataset can help researchers from a variety of disciplines interested in studying online social movements, protests, political extremism, and disinformation.

\end{abstract}

\section{Introduction}\label{sec:intro}

While the modern social media ecosystem is certainly dominated by a few major players, e.g., Facebook, Twitter, Reddit, etc., there are a variety of lesser known platforms with high active user bases.
One such platform is Telegram, a mobile messaging app that has broader social media style features.
Telegram is particularly interesting due to its use by social movements to disseminate information.
For example, the Hong Kong protesters made use of Telegram to organize some of their activities.
Unfortunately, not all uses of Telegram are generally positive: Telegram is also home to a vast network of right wing extremist groups, who use it to organize as well as disseminate racist and violent ideology.

In this paper, we present the Pushshift Telegram Dataset.
to the best of our knowledge, our dataset represents, by far, the largest collection of Telegram data made available to the public.
While our dataset is available for download as static snapshots, it is also under periodic collection. The most current snapshot is available at \url{https://zenodo.org/record/3607497}.
At the time of this writing, the Pushshift Telegram Dataset comprises 27,801 \emph{channels} and 317,224,715 \emph{messages} from 2,200,040 unique \emph{users}.

Also, we make publicly our data collection source code at \url{https://github.com/pushshift/telegram}. 
The code can be re-used by other researchers that want to take information from specific Telegram channels for their research.

In the remainder of this paper, we provide some background on Telegram, describe our dataset, and perform some general characterization of the dataset.

\section{Background \& Related Work}\label{sec:background}

\descr{Telegram} was started in August 2013 as an encrypted instant messaging platform.
Telegram users provide a telephone number to access the service. 
Messages between users are stored in a centralized cloud-based storage and (with the exception of ``secret chats'') can be accessed by multiple devices that have been linked to a single user's account.
In addition to the user-to-user secure messaging features, Telegram added broadcast \emph{channels} for one-to-many communication in September 2015. 
Such broadcast channels can be created by any Telegram user, and other Telegram users can join or subscribe to the channel to read its content.
Content in a channel mostly consists of messages, which take the form of text, still images, audio files, video files, and so on.
Other messages sent in channels are \emph{service messages}, for example status messages or errors, but these are largely invisible to regular users of the platform.
These broadly-available and widely used Telegram channels are designed for information dissemination thus are the subject of this dataset project.

\descr{Previous work that used Telegram.}
A large body of previous work studies the Telegram ecosystem itself, or uses data from Telegram to study specific emerging research problems.
Specifically,Anglano et al.~\cite{anglano2017forensic} and Satrya et al.~\cite{satrya2016digital} study the artifacts generated by the Telegram android application on the Android platform.
They propose a methodology that enable the reconstruction of important information involved in the Telegram application like list of contacts, messages exchanged between users, and information about the channels and groups that the user is involved.

Sutikno et al.~\cite{sutikno2016whatsapp} study the features that are available in three popular messaging applications: WhatsApp, Viber, and Telegram.
They conclude that Telegram is the best messaging applications in terms of security, WhatsApp is best in terms of ease-of-use, while Viber is another good option with many integrated features like in-app voice calls.
Abu-Salma et al.~\cite{abu2017security} undertake a user study to understand whether end-users understand the security features that Telegram offers. 
They find that most of the users tend to use less secure features of the Telegram application, and they overall feel secure because they ``are using a secure tool.''
Also, the authors analyze Telegram's user interface (UI) and find that it includes a lot of technical jargon and inconsistencies, and that some of the security features offered by the platform are not explained clearly in the UI.

Nikkah et al.~\cite{nikkah2018telegram} study the use of Telegram by Iranian immigrants during their immigration procedure by observing 30 Iranian immigration-related groups.
They show several examples of how Telegram bots are used enforce specific policies, how groups are moderated, and how pinned messages are posted by administrators.
Hashemi and Chahooki~\cite{hashemi2019telegram} study the group features of 900K Iranian channels and 300K Iranian groups on Telegram with the goal to identify high-quality groups (e.g., professional and business groups) over low-quality groups (e.g., dating groups).
They find that high-quality groups tend to have more phone numbers in their messages, have longer messages, and have more user engagement when compared to low-quality groups.
Asnafi et al.~\cite{asnafi2017using} examine the use of the Telegram platform in various Iranian academic libraries. 
They collect data from channels posted on the websites of the libraries and find that users mostly talked about news and information, book introductions, and various files. 
They also note that most of the messages contained images.
Akbari and Gabdulhakov~\cite{akbari2019platform} study the ban of Telegram by Russia and Iran following Telegram's refusal to give access to encrypted data on the platform.

Dargahi Nobari et al.~\cite{dargahi2017analysis} perform a structural and topical analysis of content posted on the Telegram platform. 
By collecting data from 2.6K groups/channels and 219K messages, they build a graph based on the mentions, concluding that the mentions graph is extremely sparse and includes several separated connected components that indicates that users have a low tendency to mention other users. 

Other previous research focuses on studying how terrorist organizations like ISIS use Telegram for various purposes like dissemination of content and ideology, as well as recruiting fighters and terrorists~\cite{prucha2016and,yayla2017telegram,shehabat2017encrypted}.

\descr{Other dataset papers.} Since the main contribution of this paper is the rich dataset we release, here we briefly overview previous work that focuses on releasing datasets from various Web communities.
Garimella and Tyson~\cite{garimella2018whatapp} study the WhatsApp messaging platform and they share their tools for obtaining public WhatsApp groups data, as well as a dataset from 178 groups that includes data for 454K messages posted from 45K users. 

Fair and Wesslen~\cite{fair2019shouting} focus on Gab by releasing a dataset that includes 37M posts and 24M comments posted between August 2016 and December 2018.
Zignanie et al.~\cite{zignani2019mastodon} focus on Mastodon, a decentralized social network, by releasing a dataset of 5M posts: each post is associated with a label indicating whether the post or its content is inappropriate (according to the users that made the post). 
Founta et al.~\cite{founta2018crowdsourcing} provide a large scale dataset of tweets that are annotated on whether they are hateful, abusive, spam, or a normal tweet.
To do this, they leverage crowd workers and annotate each tweet according to the majority agreement between all the crowd workers.
Brena et al.~\cite{brena2019news} release a data collection pipeline and a large scale dataset related to the dissemination of news articles on Twitter.
The data collection relies on a list of news sources and generates a large dataset of articles from these sources that are posted on Twitter.
Salem et al.~\cite{salem2019fa} focus on the Syrian War and release a carefully curated dataset of 804 news articles that are also labeled as real or fake.
Norregaard et al.~\cite{norregaard2019nela} release a set of 713K news articles collected between February and November, 2018, from 194 news sources.
Also, for each news source in their dataset, they include ratings from eight different assessment sites that include, among others, scores related to the reliability, trust, bias, and journalistic standards of each news source. 

\section{Description of the Pushshift Telegram Dataset}

\begin{figure*}[t!]
\centering
\includegraphics[width=0.8\textwidth]{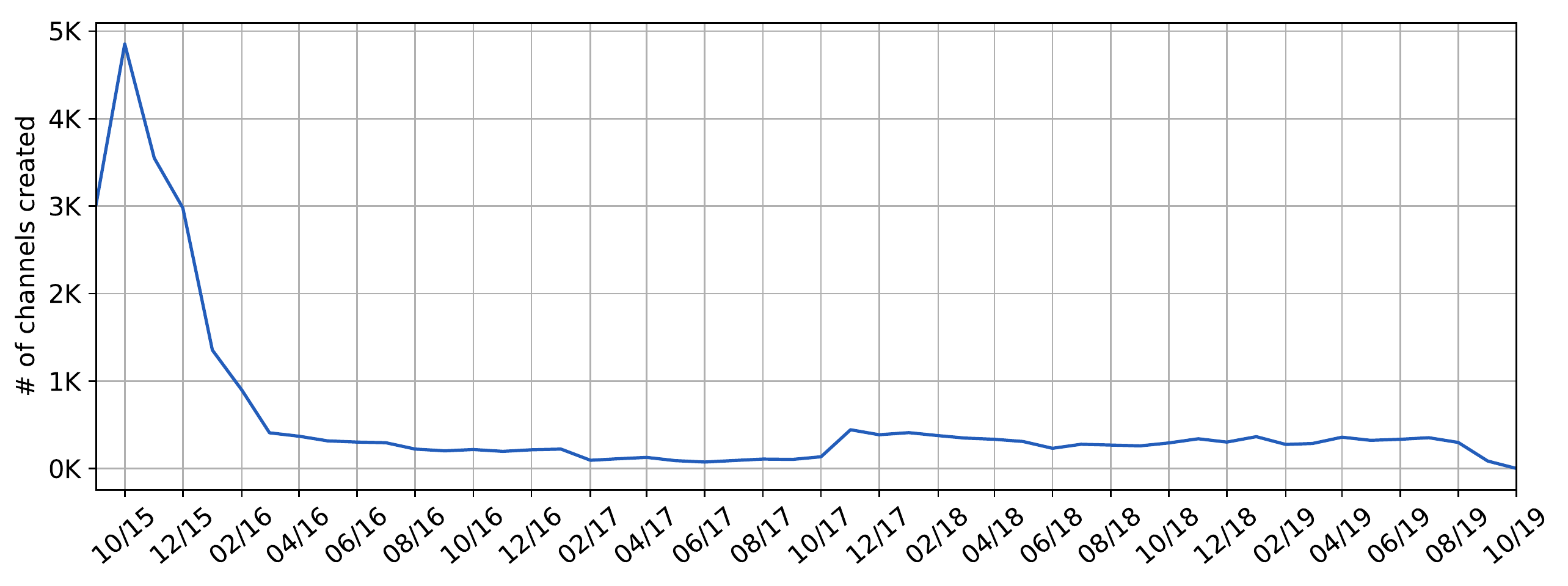}
\caption{Number of channels created per month.}
\label{fig:temporal_channel_creation}
\end{figure*}

\subsection{Data Collection}

Our data collection on Telegram is channel-based.
Our goal is to collect data and metadata for publicly-viewable channels and public ``chats''.
Typically each Telegram channel is set up by its owner to allow broadcast, or one-way, communication from a small set senders to a broader set of general channel users.
However, each Telegram channel can also optionally have an associated ``chat'' that allows communication among all channel participants.
Finally, some non-chat channels are not broadcast-only, but have themselves been configured to allow two-way communication between all participants on the channel.

To collect content and metadata from all of these types of channels and chats, we use Telethon\footnote{\url{https://docs.telethon.dev/en/latest/}}, a Python interface to the Telegram API.
We began with a seed list of approximately 250 primarily English-language broadcast channels and chat channels on Telegram.
This initial list consists of 124 channels focusing on right-wing extremist politics and 137 channels cryptocurrency-related channels.  
To grow the list of channels, we rely on the ``forwarding'' feature within Telegram wherein content can be forwarded between channels.
Each time we discover content forwarded from a channel that is not already on our list, we add it, collect its data, and follow all of its channels.
This snowball ``crawling'' approach has so far resulted in a list of 27,801 channels.

Channel data exposed by the Telegram API includes metadata such as the unique identification number, title, creation date, and various channel settings (e.g. usage configurations, administrator restrictions, and whether the channel is a bot), as well as the actual messages sent in the channel. 
Some calculated fields are also included, such as a current count of users, the identification number of the current ``pinned'' post, and a count of how many messages are unread.

Users communicate on Telegram channels by sending ``messages'' to the channel. 
Messages can either be original content posted to a channel, or can be forwarded content from another channel or from another user. 
The access rights to read a message can be restricted by the channel settings, for example allowing anyone to read the messages, or requiring users to ``join'' the channel before being able to view the messages. 
Message data exposed by the Telegram API includes metadata such as the datetime that the message was sent, whether it included media (e.g. images or video), and the identification number of the user who sent the message. 
For each message sender, Telegram provides details such as their username, whether they are a bot, whether they are a verified user, and so on. 

We stored data and metadata for each channel/chat, message, and user in a PostgreSQL relational database management system.
We currently do not collect media that accompanies the messages. 
This will change in the future as we find a more robust storage space solution.

\subsection{Dataset Structure} 
Our static snapshot consists of three files:
\begin{itemize}
	\item \emph{Accounts metadata:} A newline delimited JSON file that includes the metadata for all the accounts that posted on any of the channels in our dataset (2.2M). Documentation for the fields included in this file are provided from the Telethon library and is available at \url{https://tl.telethon.dev/constructors/user.html}.
	\item \emph{Channels metadata:} A newline delimited JSON file that includes the metadata for all the channels in our dataset (27.8K). The documentation for the fields included in the channels metadata file are available at \url{https://tl.telethon.dev/types/chat_full.html} and \url{https://tl.telethon.dev/constructors/channel_full.html}.
	\item \emph{Messages:} A newline delimited JSON file including all the messages posted in these channels (317M). A documentation of the fields included in this file are provided by the Telethon library and is publicly available at \url{https://tl.telethon.dev/constructors/message.html}
\end{itemize}

\subsection{FAIR principles}

We emphasize that our dataset fully conforms with the FAIR principles.\footnote{\url{https://www.go-fair.org/fair-principles/}}
Specifically, our dataset is \emph{Findable} since it is publicly available via the Zenodo service\footnote{\url{https://zenodo.org/}}, which assigns a digital object identifier (DOI): \href{https://zenodo.org/record/3607497}{10.5281/zenodo.3607497}.
Our dataset is also \emph{Accessible} since it can be accessed by anyone in the world, while at the same time the format of the dataset is JSON, which is a widely used standard for data format.
Due to the use of the widely known JSON standard, our dataset is \emph{Interoperable} as almost every programming language has a library to work with data in JSON format.
Finally, we release the full dataset and we provide a description of the dataset and the pointers to the Telethon API documentation that allow the interested researchers to understand the data and work with it. We ask that researchers cite our work if they use the dataset. Thus, our dataset is \emph{Reusable}.

\begin{figure}[t!]
\centering
\includegraphics[width=0.85\columnwidth]{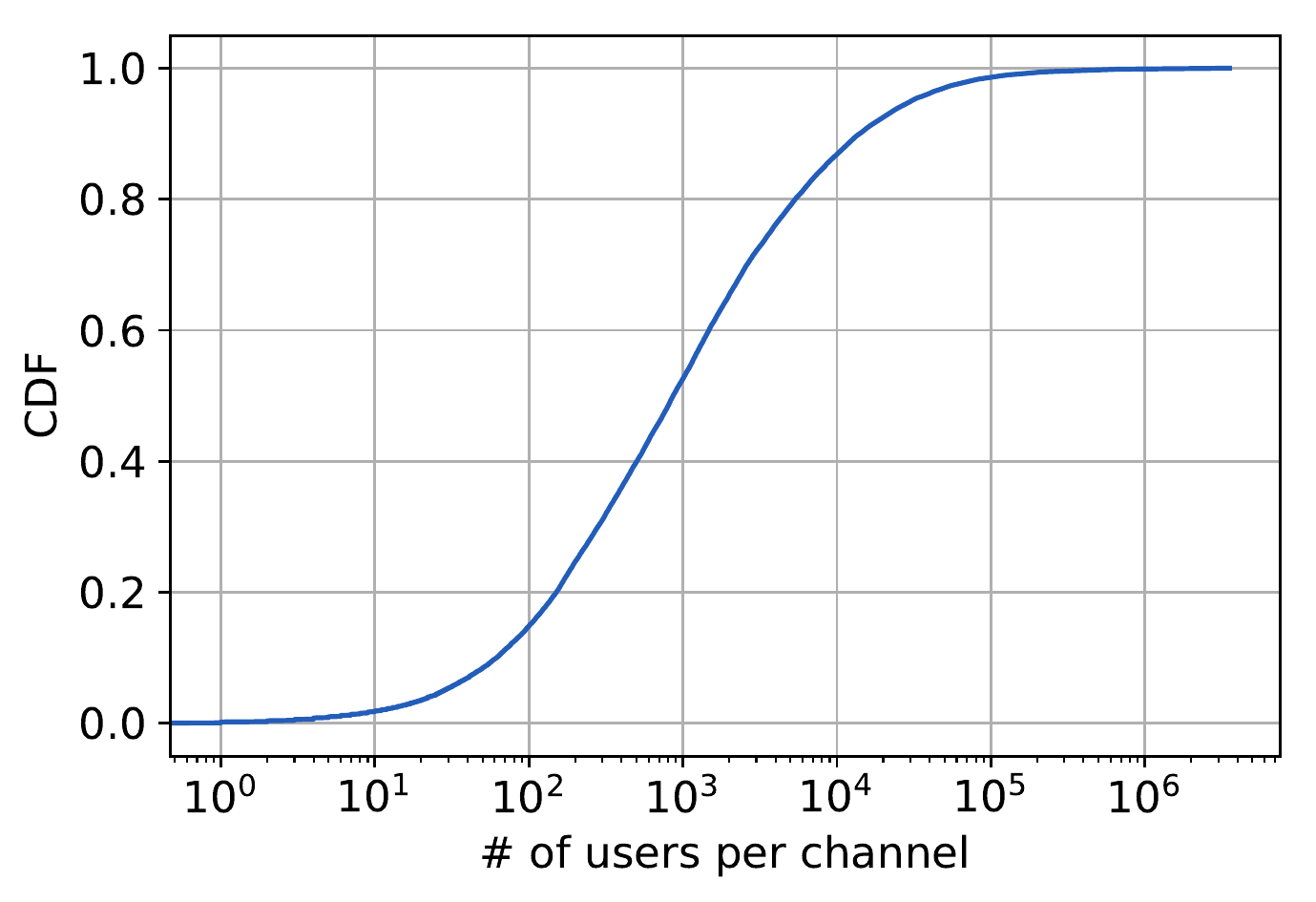}
\caption{CDF of the number of registered users per channel.}
\label{fig:cdf_users_per_channel}
\end{figure}

\begin{figure}[t!]
\centering
\includegraphics[width=0.85\columnwidth]{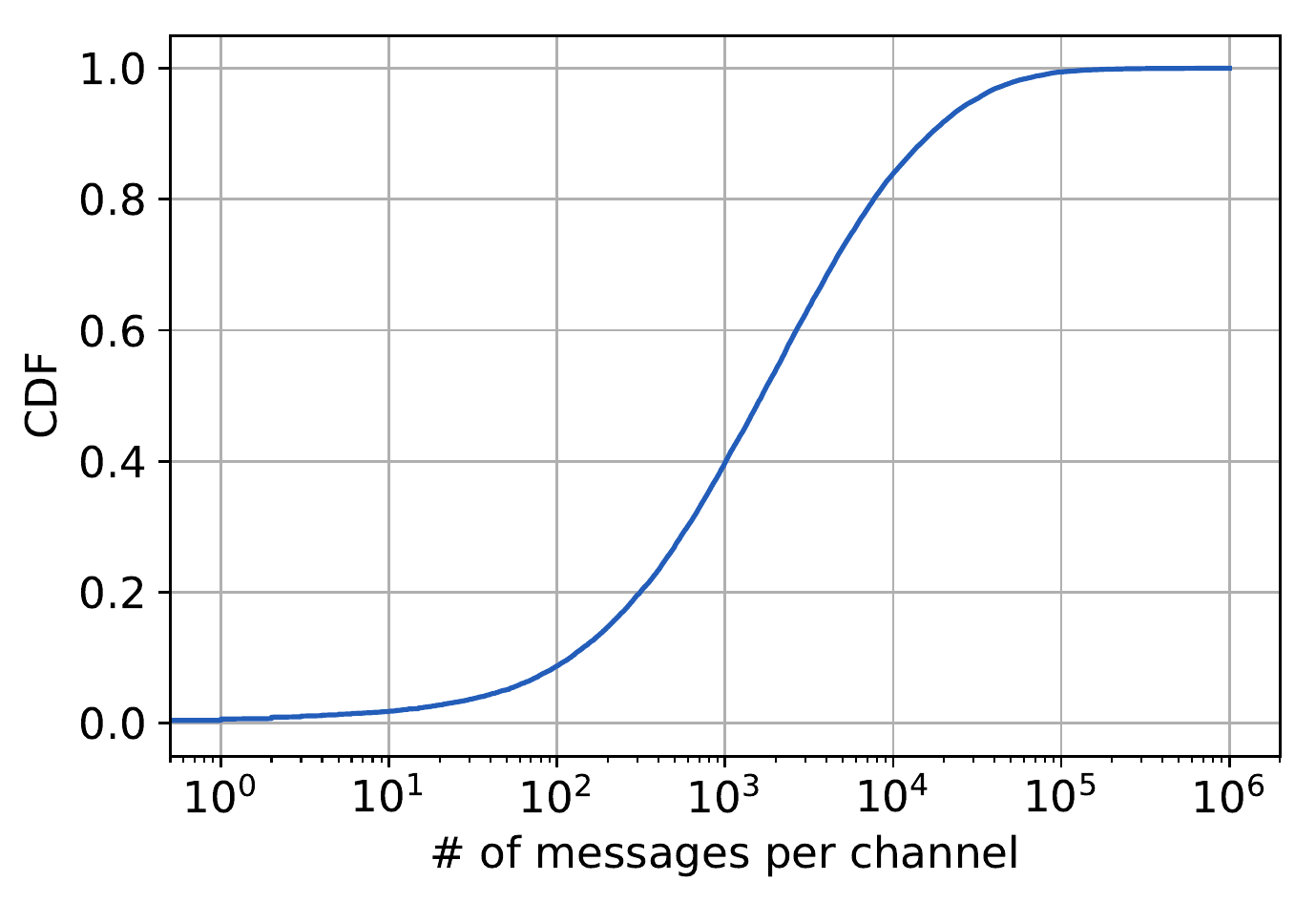}
\caption{CDF of the number of messages per channel.}
\label{fig:cdf_messages_per_channel}
\end{figure}

\begin{figure*}[t!]
\centering
\includegraphics[width=0.8\textwidth]{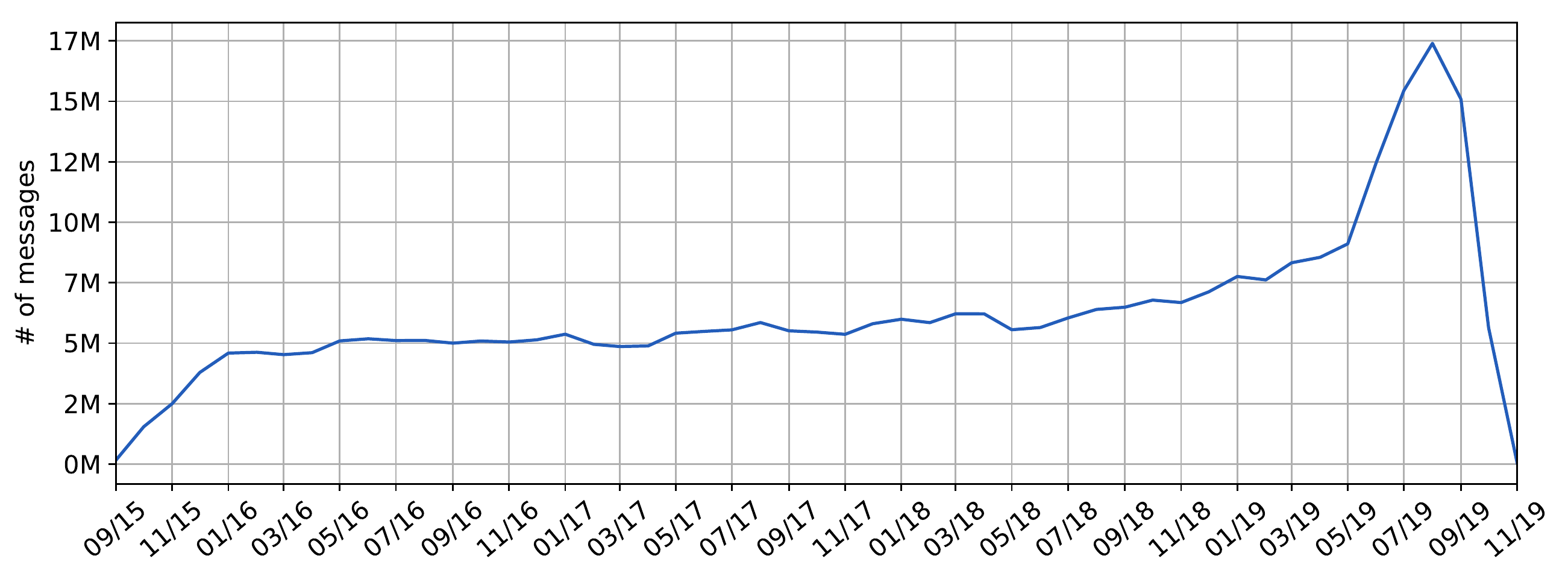}
\caption{Temporal overview of the messages that are included in our dataset. We show the number of messages per month.}
\label{fig:temporal_messages}
\end{figure*}

\section{Dataset General Characterization}

\descr{Channels.} Overall, our dataset includes information from 27,801 channels. 
Fig.~\ref{fig:temporal_channel_creation} shows how these channels are created over time: we plot the number of channels that are created per month. 
Telegram introduced channels as a feature in September 2015, and our data includes 3,024 channels created in that very first month.
We find that the channel creations spike on October 2015 with 4,854 channels created.
New channel creations drop in subsequent months.
Between March 2016-October 2017, and November 2017-August 2019, we observe a steady rate of channel creation, with the latter period having slightly more channels created.

Next, we look into the number of registered users per channel in our dataset. 
We find a mean number of registered users of 9823.3, while the median is 864.
Fig.~\ref{fig:cdf_users_per_channel} shows the Cumulative Distribution Function (CDF) of the number of registered accounts per channel. 
We observe that the majority of the channels in our dataset (85.2\%) have at least 100 registered users, while 47.4\% of the channels have at least 1000 registered users.

Fig.~\ref{fig:cdf_messages_per_channel} shows the number of messages per channel. 
We find, a mean of 6937.5 messages per channel, while the median is 1644.
Also, we observe that 60\% of the channels have at least 1K messages.

\begin{figure}[t!]
\centering
\includegraphics[width=\columnwidth]{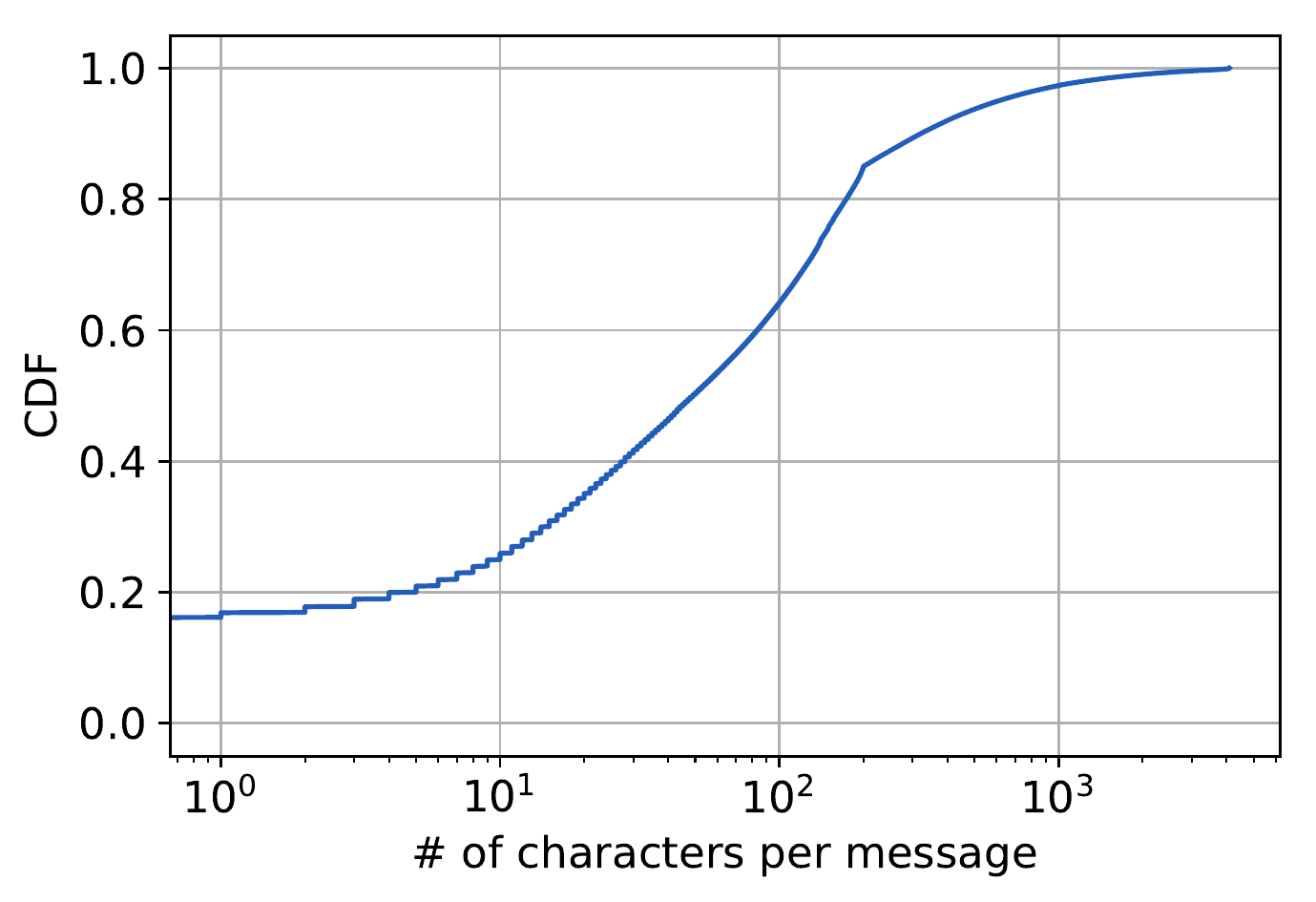}
\caption{CDF of the number of characters per message.}
\label{fig:cdf_messages_length}
\end{figure}

\descr{Messages.} Overall, our dataset includes information for 317,224,715 messages. 
Out of those, 3,069,829 are just service messages indicating an event that happened on a specific channel (e.g., user adds, errors, and other status messages). 
The remaining 314,154,886 messages are actual messages that include content shared on the channel by users. 
Fig.~\ref{fig:temporal_messages} shows the monthly number of non-status messages that are shared in our dataset.
We observe that, during 2016 and 2017, message activity is somewhat stable with around 5M messages per month, while we find a peak in message activity during August 2019 with approximately 17M messages.
This peak in August 2019 coincides with the addition of 19,000 new Telegram users during the Hong Kong protests-\cite{hongkong}.

Fig.~\ref{fig:cdf_messages_length} shows the CDF of the number of characters per message, which gives an intuition of how lengthy Telegram messages are. 
We find that messages have a mean number of characters equal to 152.2, while the median is 49 characters per message.
We also observe that a substantial percentage of messages (16.1\%) have an empty message, likely indicating that users are sharing messages with multimedia and no textual message.

\begin{figure}[t!]
\centering
\includegraphics[width=\columnwidth]{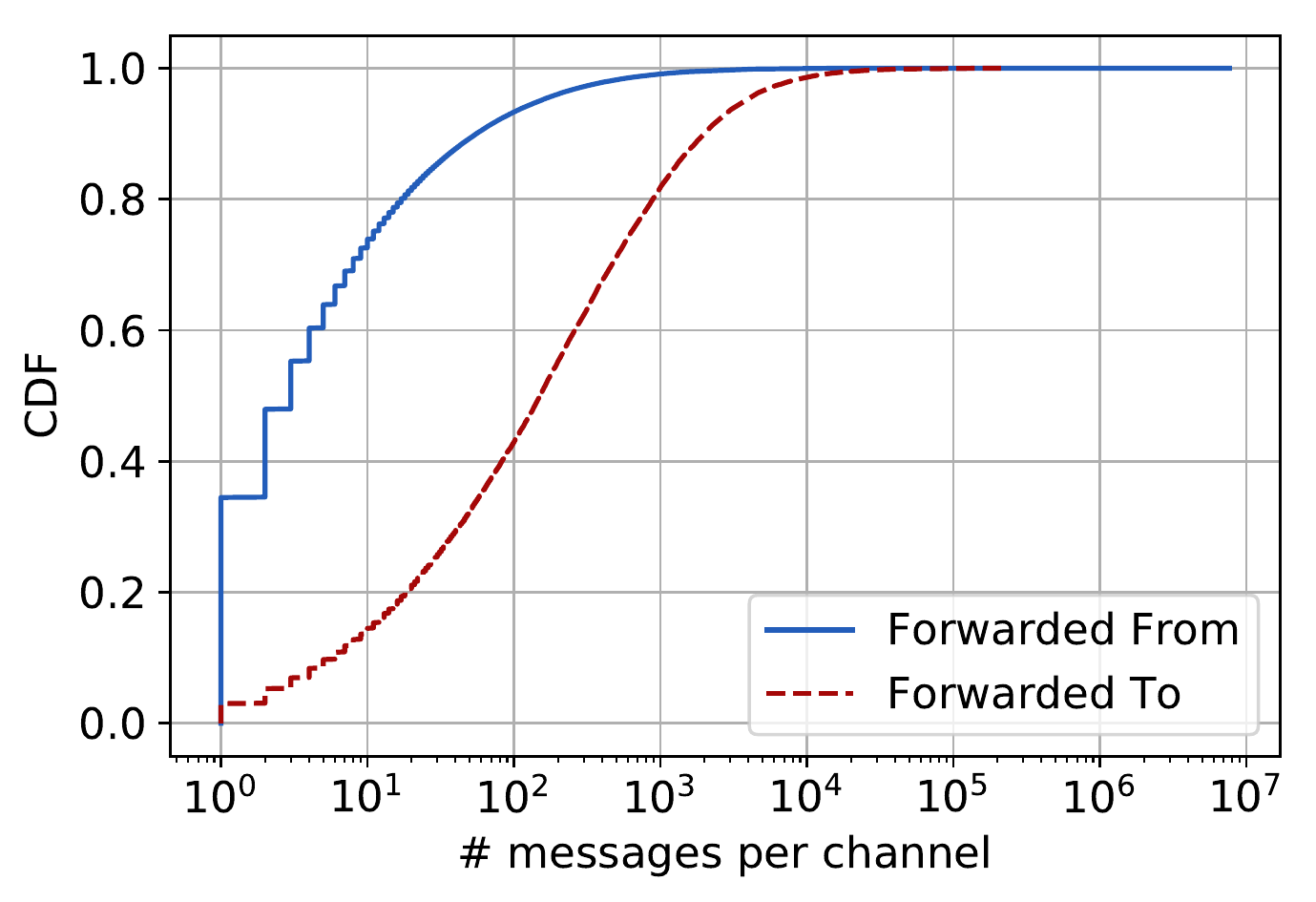}
\caption{Distribution of number of messages per channel that were forwarded.}
\label{fig:message_forwards}
\end{figure}

\descr{Forwarded Messages.} We also assess how common it is to forward messages across channels on Telegram. 
Other popular messaging apps like WhatsApp limit the number of times a specific user can forward a specific message, in an attempt to limit the spread of misinformation~\cite{whatsapp_forwards}.
Previous research suggest that this counter measure can offer substantial delays in the propagation of misinformation~\cite{de2019can}.
Telegram does not limit the number of times a message can be forwarded.
In our dataset, we find that 25,601,073 (8.1\%) of all the messages are actually forwards from previously posted messages on other channels, or forwards from messages other users posted to their personal user channels.
This indicates that forwarding messages across Telegram channels and between users and broadcast channels is a common operation on the platform.
Next we investigate the users and broadcast channels involved in the forwarding of messages. 
We find 346,937 user channels or broadcast channels where forwarded messages originate. 
At the same time we find 27,039 broadcast channels where messages were forwarded to.
Fig.~\ref{fig:message_forwards} shows the CDF of the number of messages per channel for the the channels that messages were forwarded from and to.
We observe that the number of messages-per-channel in the channels forwarded \emph{from} is substantially smaller than the number of messages-per-channel in the channels forwarded \emph{to}. 
The \emph{from} mean is 946.8 vs \emph{to} mean of 73.9, while \emph{from} median is 151 vs \emph{to} median of 3.
This difference is due to the number of channels that messages were forwarded from is substantially larger (346K vs 27K).
Overall, these findings indicate that message forwarding across Telegram channels is a popular feature of the Telegram platform, and this feature can be studied by researchers to assess the effect of this feature in emerging phenomena like the spread of false information, hateful content, etc.

\begin{figure}[t!]
\centering
\includegraphics[width=\columnwidth]{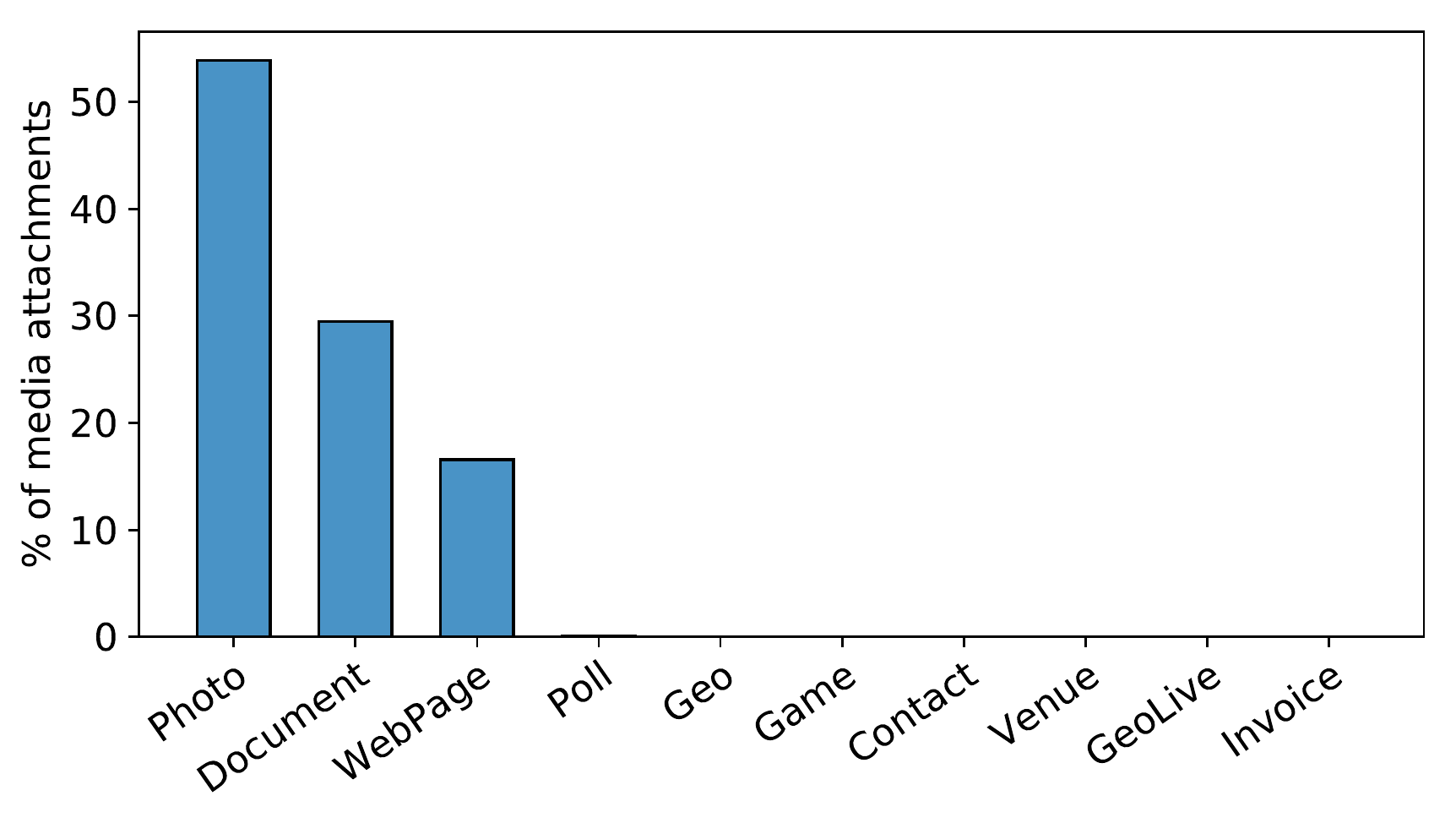}
\caption{Distribution of the media attachments according to their type.}
\label{fig:media_types}
\end{figure}

\begin{figure}[t!]
\centering
\includegraphics[width=\columnwidth]{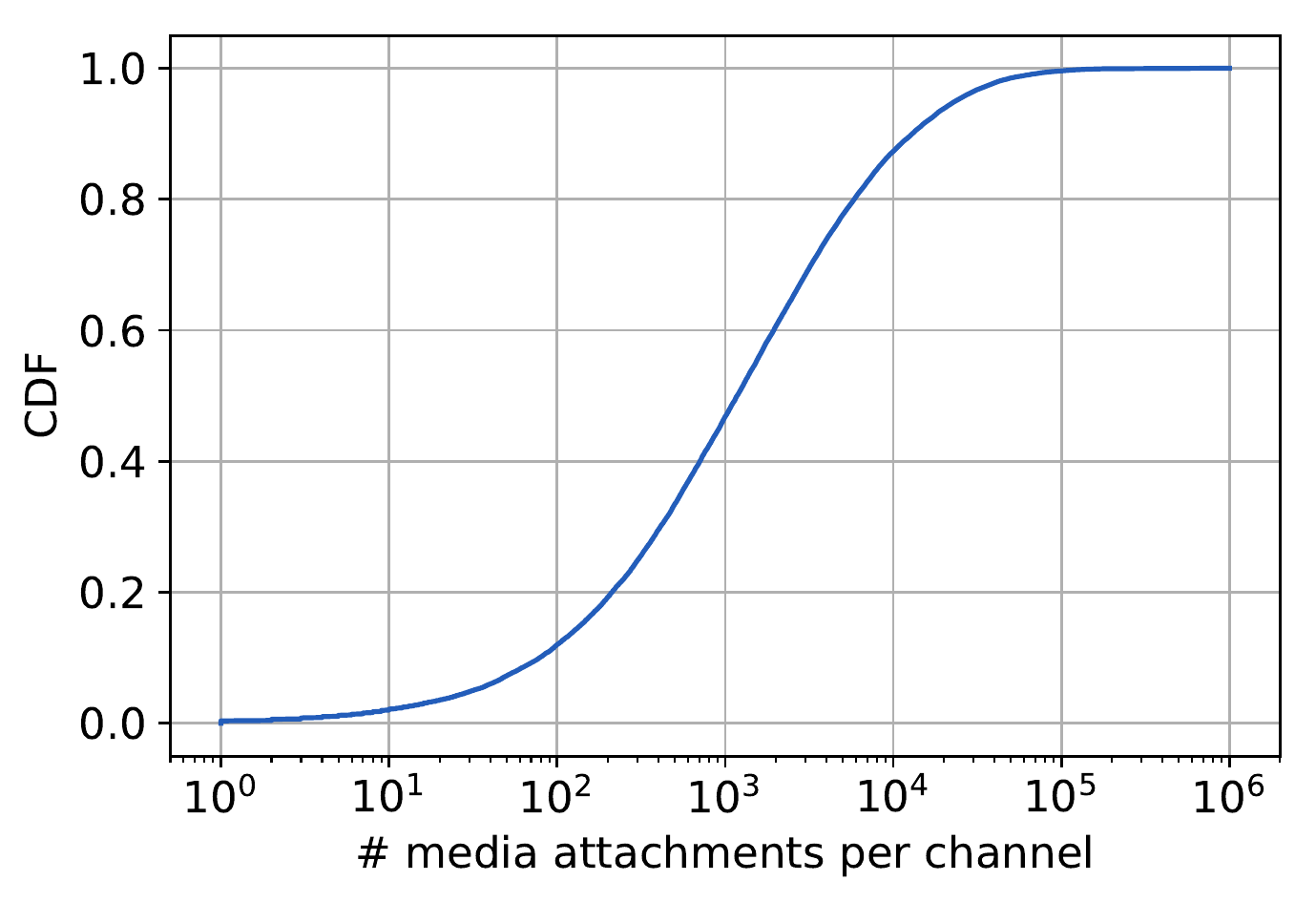}
\caption{CDF of the number of media attachments posted per channel}
\label{fig:cdf_media_per_channel}
\end{figure}

\descr{Media.} Next we look into whether messages contain media attachments, and what the different types of media attachments are in our dataset.
We observe that 50.1\% of the messages include media attachments with 48.1\% of all the messages containing exactly one attachment.
Fig~\ref{fig:media_types} shows the distribution of all the media attachments into types according to Telegram. 
We observe that in our dataset, 53.8\% of the attachments are photos, 29.4\% are documents, 16.5\% are Web pages, while the rest 0.3\% of the attachments are Polls, Geo locations, Games, Contacts, Venues, or Invoices. 
Fig.~\ref{fig:cdf_media_per_channel} shows the CDF of the number of media per channel in our dataset.
We observe that Telegram is a rich source of media: more than half of the channels have shared over 1K media attachments.

\descr{Mentions and Hashtags.} Here we investigate whether Telegram users are using mentions and hashtags in their messages.
Overall, we find 7,638,430 (2.3\%) messages contain hashtags, and 87,029,573 (27.7\%) messages contain mentions.
This indicates that Telegram users are more frequently mentioning other users or channels when compared to using hashtags in their messages.
Table~\ref{tbl:top_mentions_hashtags} report the 20 most popular mentions and hashtags that we find in our dataset, as well as their respective percentage over the overall number of messages that include hashtags/mentions.
In terms of hashtags, we observe several international hashtags like \#Trending, \#Hot, \#news. 
Interestingly, we also find divisive hashtags like \#USAKillsYemeniPeople.
In terms of mentions, we observe that most of them is related to Iran, hence highlighting the popularity of Iranian users on Telegram in general and in particular in our collected sample.

\begin{figure}[t!]
\centering
\includegraphics[width=\columnwidth]{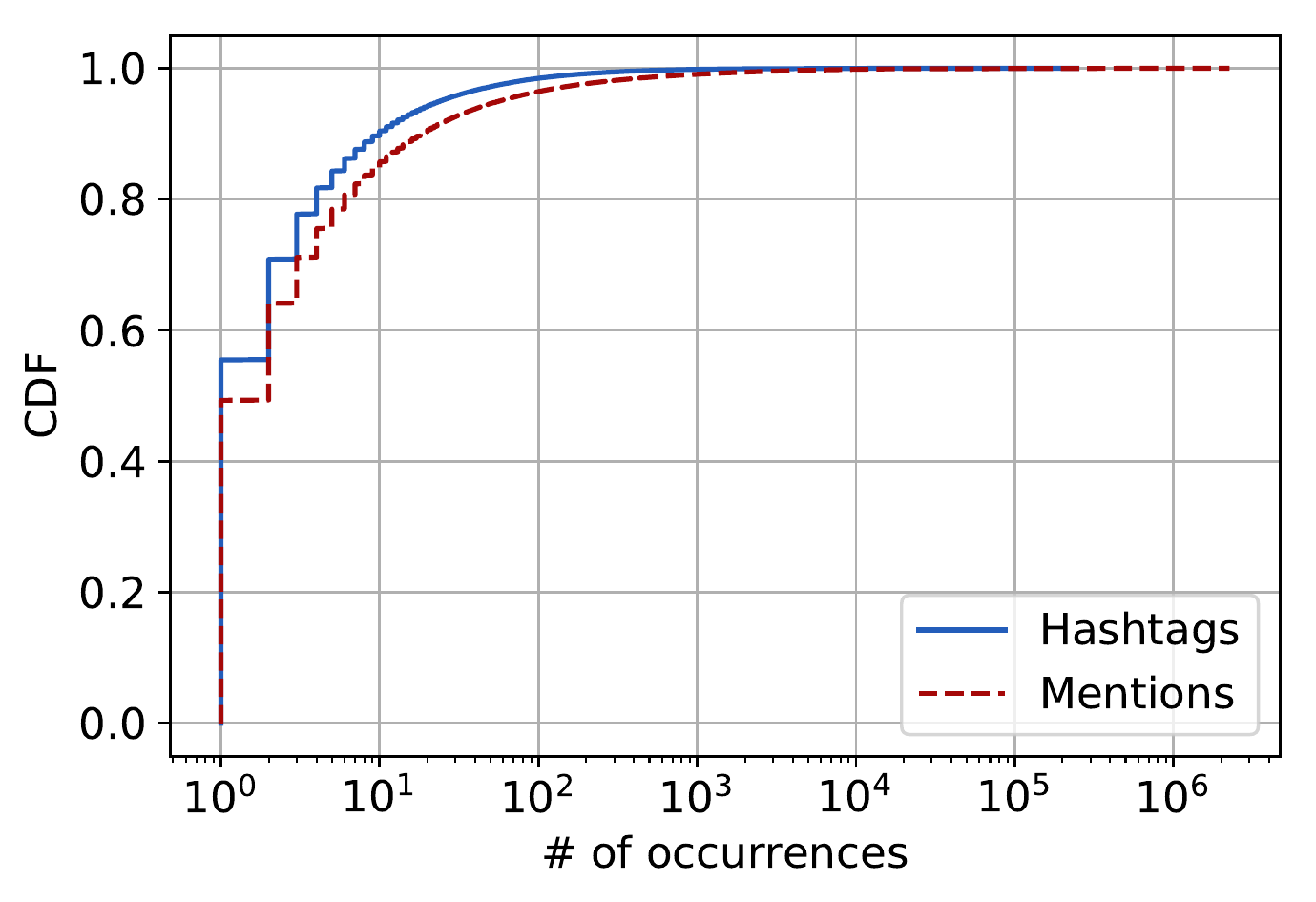}
\caption{CDF of the number of occurrences per hashtag/mention.}
\label{fig:cdf_occurrences_hashtags_mentions}
\end{figure}

We also plot the number of occurrences per hashtag/mention in Fig.~\ref{fig:cdf_occurrences_hashtags_mentions}.
For hashtags we find a mean number of 14.3 occurrences per hashtag, while 55.4\% of the hashtags occur only once in our dataset.
For mentions we find a mean number of 82.4 occurrences per mention, while 49.3\% of the mentions occur only once in our dataset.

\begin{table}[]
\centering
\resizebox{\columnwidth}{!}{%
\begin{tabular}{@{}lrlr@{}}
\toprule
\textbf{Hashtag}     & \multicolumn{1}{l}{\textbf{\begin{tabular}[c]{@{}c@{}} \% \\(out of 7.6M)\end{tabular}}} & \textbf{Mention}      & \multicolumn{1}{c}{\textbf{\begin{tabular}[c]{@{}c@{}}\% \\(out of 87M)\end{tabular}}} \\ \midrule
Trending             & \multicolumn{1}{r|}{3.23\%}                                                                          & tahlilgarantala       & 2.59\%                                                                                              \\
Hot                  & \multicolumn{1}{r|}{2.17\%}                                                                          & tahlilgarantala\_ons  & 1.15\%                                                                                              \\
Syria                & \multicolumn{1}{r|}{1.07\%}                                                                          & tahlilgarantala\_absh & 0.89\%                                                                                              \\
ULTIMORA             & \multicolumn{1}{r|}{0.81\%}                                                                          & noticiasul            & 0.59\%                                                                                              \\
request              & \multicolumn{1}{r|}{0.75\%}                                                                          & tahlilgarantala\_Seke & 0.55\%                                                                                              \\
Step\_News           & \multicolumn{1}{r|}{0.64\%}                                                                          & Twitter\_Farsi        & 0.48\%                                                                                              \\
Nima                 & \multicolumn{1}{r|}{0.55\%}                                                                          & ilnair                & 0.27\%                                                                                              \\
habrahabr            & \multicolumn{1}{r|}{0.55\%}                                                                          & TEQNYEBOT\_BOT        & 0.27\%                                                                                              \\
Geral                & \multicolumn{1}{r|}{0.51\%}                                                                          & MyAsriran             & 0.23\%                                                                                              \\
USAKillsYemeniPeople & \multicolumn{1}{r|}{0.48\%}                                                                          & iran\_times           & 0.21\%                                                                                              \\
Mundo                & \multicolumn{1}{r|}{0.44\%}                                                                          & haberbulteni          & 0.20\%                                                                                              \\
Venezuela            & \multicolumn{1}{r|}{0.42\%}                                                                          & khabaredagh           & 0.19\%                                                                                              \\
Amazon               & \multicolumn{1}{r|}{0.42\%}                                                                          & BI20ST                & 0.19\%                                                                                              \\
news                 & \multicolumn{1}{r|}{0.41\%}                                                                          & Farsna                & 0.18\%                                                                                              \\
Marvel               & \multicolumn{1}{r|}{0.39\%}                                                                          & K\_BER1               & 0.18\%                                                                                              \\
Sport                & \multicolumn{1}{r|}{0.38\%}                                                                          & KNWAT                 & 0.15\%                                                                                              \\
tw                   & \multicolumn{1}{r|}{0.35\%}                                                                          & T5TTI                 & 0.15\%                                                                                              \\
soc                  & \multicolumn{1}{r|}{0.34\%}                                                                          & caspiankhabar         & 0.14\%                                                                                              \\
Economia             & \multicolumn{1}{r|}{0.34\%}                                                                          & alalamnewstv          & 0.14\%                                                                                              \\
SouCurioso           & \multicolumn{1}{r|}{0.33\%}                                                                          & Khabar\_Varzeshi      & 0.14\%                                                                                              \\ \bottomrule
\end{tabular}%
}
\caption{Top 20 hashtags and mentions that we find in our dataset}
\label{tbl:top_mentions_hashtags}
\end{table}

\section{Discussion \& Conclusion}\label{sec:conclusion}

In this paper, we described the Pushshift Telegram Dataset, to the best of our knowledge, the largest and most comprehensive Telegram dataset available to date.
Our dataset includes over 317M messages from 2.2M unique users across 27.8K channels.
In addition to the data, we also release the source code we used to collect it.
Our dataset can be used by researchers to advance the frontier of knowledge around a variety of topics.
For example, our dataset includes a large number of messages from right wing extremist groups, as well as more global movements like the Hong Kong protesters.
Thus, researchers interested in understanding how computer mediated communication affects and is used by disinformation campaigns, violent organization, as well as more traditional political protests will find great value in the dataset presented herein.
Along with the static snapshot, we supply the source for collecting data from Telegram channels.
We argue that this implementation will be extremely useful to researchers that are interested in studying Telegram and its various aspects.

\small
\bibliographystyle{abbrv}

\end{document}